\documentclass[aps,prb,floatfix,twocolumn,showpacs]{revtex4-1}
\usepackage{latexsym}
\usepackage{graphicx}
\newcommand{\figwidth}{3.275 in}
\usepackage{epsfig}
\usepackage{amsmath}
\usepackage{color}
\begin{document}

\title{Ab initio studies of opto-electronic excitations in VO$_2$}
\author{Adam Gali$^{(1,2)}$}\email{agali@eik.bme.hu}, 
\author{John E. Coulter$^{3}$ and  Efstratios
  Manousakis$^{(3,4)}$ }  
\affiliation{$^{(1)}$Institute for Solid State Physics and Optics, Wigner
  Research Center for Physics,\\
  Hungarian Academy of Sciences, Budapest, P.O.B. 49, H-1525 \\
$^{(2)}$Department of Atomic Physics, Budapest
              University of Technology and Economics, Budafoki \'ut 8., H-1111,
              Budapest, Hungary \\
$^{(3)}$ Department  of  Physics and National High Magnetic Field Laboratory,
  Florida  State  University,  Tallahassee,  FL  32306-4350,  USA\\
$^{(4)}$Department   of    Physics,   University    of   Athens,
  Panepistimioupolis, Zografos, 157 84 Athens, Greece.}

\date{\today}

\begin{abstract}
We study the optical response of VO$_2$ in the M1 insulating phase
using methods based on density functional theory in its most recent
developments. We start from a hybrid functional approach which may be
a good starting point to carry out many-body perturbation theory
since, as we show, it gives a qualitatively and to some degree
quantitatively correct description of the insulating phase.  In order
to calculate the dielectric function, first, using hybrid density functional
wavefunctions we have added GW corrections on top of the hybrid
density functional calculations and we solved the so-called Bethe
Salpeter equation to include effects of correlations of the
electron-hole pair created upon photon absorption. We find that the
effects of electron correlations are very important and they show up
as a strong contribution of the electron-hole interaction in calculating
the effects of the Bethe-Salpeter equation in the electron/hole pair
excitation on top of the interacting group state.
\end{abstract}
\maketitle

\section{Introduction}

A plethora of fascinating new phenomena has been recently reported on
oxide hetero-structures of transition metal oxides (TMO) and
devices.\cite{Mannhart, Cen, Gozar} For example, an interface between
two insulators behaves as a metal \cite{Cen} which becomes
superconducting at sufficiently low temperatures, while an interface
between two antiferromagnets becomes ferromagnetic.\cite{Millis}
These new structures not only create a playground for unexpected
physical phenomena to be observed, but, in addition, they open up the
possibility for new applications based on radically different
foundations.  Probing such interesting new phenomena is quite difficult
due to impurity effects, lattice imperfections, and other
materials-growth limitations.  Very recent progress in material
synthesis, \cite{Mannhart, Gozar, Bhattacharya, Adamo, Bhattacharya2,
  Warusawithana, Warusawithana2} however, is starting to make it
possible to carefully investigate these fascinating systems.

The complex, unusual, and as yet not fully discovered or understood
behavior of these strongly correlated materials, such as transition
metal oxides (TMO), can be manipulated in a variety of fundamentally
new applications. In particular, we would like to mention one such
possibility which has been recently proposed \cite{Mottsolar} by one
of the authors of the present paper.  This is based on the fact that
these strongly correlated localized electrons form an electronic
system which can be near the metal to Mott-insulator
transition.\cite{Mott} Photovoltaic devices based on carefully chosen
doped Mott insulators can produce a significant photovoltaic
effect. More importantly, it was found \cite{Mottsolar} that if the
Mott insulator is appropriately chosen, the photovoltaic effect can
lead to solar cells of high efficiency, where a single solar photon
can produce multiple electron/hole pairs, an effect known as impact
ionization, in a time-scale shorter than the time characterizing other
relaxation processes. Increase of solar cell efficiency due to this
process has been proposed previously for band-gap semiconductors;
however, the effect is not significant there, because the
characteristic time-scale for impact ionization is comparable to the
time-scale for electron-phonon relaxation.  The reason that this is
not expected to hold for a Mott insulator is that the large Coulomb
repulsion present in a Mott insulator leads to a large enhancement of
the impact ionization rate.

It is very important to have a computational \emph{ab initio} scheme
which is reliable to evaluate opto-electronic properties such as
excitations, gaps, absorption and, ideally to be able to compute the
induced photo-current for a given bulk or interface
structure. Furthermore, this scheme should be reliable for transition
metal oxides where the electronic Coulomb correlations are expected to
play an important role.  Assuming that such a scheme exists, it would
be very valuable in a variety of applications in real materials,
including in the case of photovoltaic applications proposed in
Ref.~\onlinecite{Mottsolar}.  This tool would meaningfully aid
experimental efforts towards finding the most suitable TMO based
material with suitable size gap, band structure, size of transition
matrix elements including selection rules, etc.

In this paper we choose to study the optical properties of the
insulating M1 phase of VO$_2$. Vanadium dioxide is regarded as the
prototypical example of a strongly correlated TMO material with a long
history starting with the original suggestion by Mott himself
\cite{MottVO2} and his collaborator Zylbersztejn nearly four decades
ago. Furthermore, this material has been the playground for almost all
many-body techniques available to tackle strong Coulomb correlations.

VO$_2$ undergoes a structural distortion below approximately $T_S \sim
340 K$ at ambient pressure \cite{Anderson, *Anderson2, Goodenough1}.
Above $T_S$ the structure of VO$_2$ is rutile-type and it is metallic
and below $T_S$ VO$_2$ is an insulator.  The insulating low
temperature phase is monoclinic and is called the M1 phase. We would like
to focus our studies on this phase because it is an insulator
that exists at room temperature which could be utilized for
photovoltaic applications.

\begin{figure}
\vskip 0.3 in
\epsfig{file=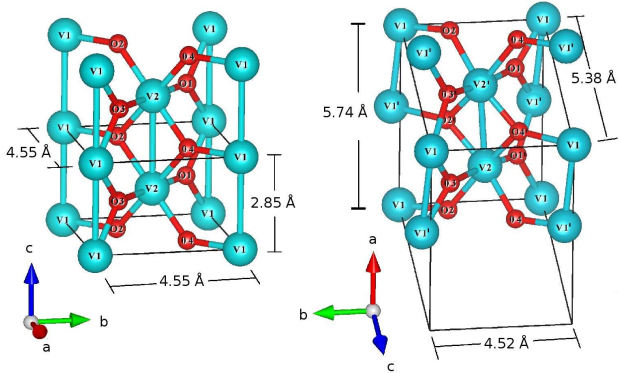,width=\figwidth}
\caption{(Color online) Unit cell of metallic and M1 phase of VO$_2$
  single crystal. The symmetry axes of these crystals are shown. The
  rutile c-axis is applied as a ``parallel'' direction for the
  distorted M1 phase in the context which is conventionally called
  a-axis in M1 phase. The experimental lattice constants are also depicted.}
\label{figure1}
\vskip 0.3 in
\end{figure}

Below $T_S$ and at room temperature, the vanadium (V) atoms dimerize
and the V-V pair tilt around the rutile c-axis, doubling the unit cell
as illustrated in Fig.~\ref{figure1}.  This dimerization causes
a zigzag-like antiferroelectric tilt of the V atoms perpendicular to
this axis. The dimerization affects mainly the V $3d$ $t_{2g}$ states,
which in the metallic phase fall near the Fermi energy and have
similar band occupations.\cite{Goodenough1, Goodenough2} In the
dimerized state, these states hybridize differently to form
$d_{\parallel}$ and $\pi^*$ states. The bonding $d_{\parallel}$ states
fall below the $\pi^*$ states, and they become fully occupied by the
single $d$ electron, leading to the insulating behavior due to this
Peierls-type distortion. Our calculations support this picture and
we illustrate what actually happens as the system undergoes the
phase transition from the metallic to the insulating M1 phase
in Sec.~\ref{PDOS} of this paper.


Previous \emph{ab initio} calculations of optical properties, such as
of the real and imaginary parts of the zero momentum dielectric
function, show vast disagreement between the result of the calculation
and the experimental results. For example, dielectric results from the
local density approximation (LDA) within density functional theory
(DFT) \cite{CA-80, PZ81} must be shifted entirely by hand to begin to
agree with experiment.\cite{Mossanek} Results from a generalized
gradient approximation of DFT (GGA) together with a
Hubbard-correction,\cite{Hubbard} GGA+$U$ look strikingly different
apart from matching the optical gap \cite{Liu} (which is effectively
set by hand by tuning $U$). In addition, the electronic structure is
predicted to be metallic rather than insulating using both LDA and
GGA.\cite{Eyert1,*Eyert2, Wentzcovitch1,*Wentzcovitch2}

Methods that go beyond ground-state DFT are now well established.
However one should remember that a better starting point is 
absolutely necessary for $d$-electrons.


With LDA+$U$ one gets antiferromagnetic insulating ground states for
both the M1 and the metallic phase.\cite{Korotin} In addition, it
treats differently the various orbitals and we do not always \emph{a priori}
know which orbitals need this special treatment. Furthermore the exact
value of the parameter $U$ is not known and it is phenomenologically
determined.

The combination of LDA and dynamical mean-field theory
(DMFT) \cite{Biermann, Tomczak} has correctly described the M1
phase. However, a parameter free theory to describe correctly both the
metallic and the insulating M1 phase is still absent.

The self-consistent COHSEX approximation of the GW-method \cite{Hedin} gives a
good description of quasi-particle states and has been applied to
VO$_2$ rather recently \cite{COHSEX} with interesting results and
conclusions.  Furthermore, several authors have used {\it hybrid
  functionals}, a rather \emph{ad hoc} procedure, and this approach
seems to be a good compromise in many systems \cite{TMOHSE} including
VO$_2$.\cite{Eyert2} There is no good theoretical justification for
the success of this procedure.

Optical absorption experiments create an interacting electron-hole
pair, the exciton.  Good agreement between theory and experiment can
only be achieved by taking into account the exciton, especially if the
system is a semiconductor or an insulator.  Small-gap semiconductors
and metals screen excitons.  For accurate absorption spectra where
such excitonic effects are important one has to solve the
Bethe-Salpeter equation (BSE) \cite{BSE} which uses the intuitive
quasiparticle picture.  The intrinsic two-particle nature of the BSE
makes the calculations very cumbersome, since a four-point equation
(due to the propagation of two particles) has to be solved.

In the present paper we go one step further in the characterization of 
the M1 phase of VO$_2$ crytal. The hybrid
functional approach provides a better starting point, with the correct
qualitative features, such as the correct symmetry of the ground state
and a finite gap for a system in the insulating state; therefore, one
expects that if we carry out many-body perturbation theory on top of a
hybrid functional approach, such as HSE06,\cite{HSE,HSE06} and we
include the perturbative corrections, i.e., HSE06+GW and HSE06+GW+BSE,
we will find a good convergence of the perturbative series.

In the following we describe our approach in the calculation of opto-electronic excitation of VO$_2$ in Section~\ref{sec:approach}. Here, we describe the applied density functionals and convergence tests on different parameters of the calculations. In the next sections we present our results and compare them to known experimental data. First, we show the calculated electronic structure with different methods and compare them to photo-emission data in Section~\ref{sec:dos}. Here, we particularly analyze the phase transition from metallic rutile structure to the M1 Mott-semiconductor structure of VO$_2$. In addition, we validate a hybrid density functional with a parameter-free many-body perturbation theory method. Next, we describe the results on the optical properties of the M1 phase of VO$_2$ in Section~\ref{sec:dielfunc}, again, and compare them with the recorded absorption spectrum. Finally, we summarize and conclude our results in Section~\ref{sec:summary}.

\section{The Approach}
\label{sec:approach}

We carried out DFT calculations on the M1 phase of VO$_2$ as
implemented within the \textsc{VASP} package.\cite{VASP, *Kresse1,
  *Kresse2} We used small core projectors for vanadium ions, so we
explicitly included $3s3p$ electrons as valence. The valence electron
states were expressed as linear combination of plane waves. We found
that the plane wave cutoff of 400~eV provided convergent single
particle levels. We used the experimental geometry in all the
calculations as shown in Fig.~\ref{figure1}. We attempted
to optimize the structure using the hybrid functional; the converged
geometry disagrees with the experimental geometry rather 
significantly (the largest disagreement of the lattice constants was 3\%), 
giving rise to a  larger gap by a factor of 2 than that obtained
by the same hybrid functional using the experimental geometry.  
We applied various
numbers of $k$-points for the unit cell depending on the calculated properties.
 We found that the required size of the Monkhorst-Pack \cite{MP76} $k$-point set depends strongly on the existence of a Mott gap.
Namely, convergent charge density could be achieved already with a 
$5\times5\times5$ Monkhorst-Pack $k$-point set when there is
a Mott-gap, while an $18\times18\times18$ $k$-point set
was required when there was no Mott-gap. 
We show below that BSE calculations needed
special treatment.

\subsection{The general methodology}

As a necessary step, first, we carry out standard LDA and GGA-type PBE
\cite{PBE} calculations for the ground state density of states (DOS),
the band structure, and the optical response (the real and imaginary
part of the dielectric function) of VO$_2$. Then, we carry out hybrid
functional calculations, such as HSE06 \cite{HSE, HSE06} which involve
using a much more computationally demanding computational scheme.  The
reason we begin from such calculations is that the standard LDA or GGA
calculations give a single particle spectrum with significant density
of states at the Fermi level (see Fig.~\ref{figure5}) and no gap. This
is a qualitatively different state from the correct ground state, and,
thus, not suitable as a starting place for a perturbative
scheme. The hybrid functionals, as we find, give a quasiparticle gap
of the same order of magnitude as the observed gap, and thus, we can
use them as a starting point in a many-body perturbation theoretical
approach to find the leading corrections.

The HSE functional for the exchange-correlation part of the energy
involves a parameter $a$ which mixes the contribution of the
short-range parts of the Hartree-Fock exchange and the PBE expression
for the exchange energy:\cite{HSE}
\begin{eqnarray}
E_{xc} = a E^{HF,SR}_{x}(\omega) + (1-a) E^{PBE,SR}_x(\omega) \nonumber \\
+ E^{PBE,LR}_x(\omega) + E_c^{PBE}
\end{eqnarray}
where the parameter $\omega$ defines what is meant by short and long
ranged part of the Coulomb potential, which is split according to
\begin{eqnarray}
\frac{1}{r} = \frac{erfc(\omega r)}{r} +
\frac{erf(\omega r)}{r},
\end{eqnarray}
where the part involving the complementary error-function 
$erfc(\omega r)=1-erf(\omega r)$ is the short-range
and the part involving the error-function itself is the long-range part.
The value of the parameter $\omega=0.2 a_0^{-1}$ is determined to give a 
balanced description that provides good accuracy and speed
for both molecules and solids.\cite{HSE06}
The $E^{HF,SR}_{x}(\omega)$ is the Hartree-Fock exchange
part which is calculated using the short-range part of the Coulomb interaction.
$E^{PBE,SR/LR}_x(\omega)$ is the Perdew \emph{et al.}\cite{PBE} 
expression for the exchange energy functional which is modified to
use the short/long range part of the Coulomb interaction.\cite{HSE}

The value of the mixing parameter $a=1/4$ has been used in a variety
of materials giving rather reasonably accurate values for the energy
gap and other quasiparticle properties. Rationale for this functional
and for the choice of this value for $a$ is given in
Ref.~\onlinecite{PEB}.  There was argued that $a=1/n$ with $n = 4$
should be the optimum choice for typical molecules for which
fourth-order M{\o}ller-Plesset perturbation (MP4) yields atomization
energies with a small mean absolute error. The case of $n>>4$ arises
when there is a nearly-degenerate ground-state of an unperturbed
problem which corresponds to the Kohn-Sham non-interacting system.  An
ideal hybrid should be sophisticated enough to optimize $n$ for each
system and property.\cite{PEB} In the present paper, for the case of
VO$_2$ we provide results for $n=4$ and for $n=8$. The former is the
so-called HSE06\cite{HSE06} functional, and the latter we will refer
to in the following as HSE-1/8. We also refer to HSE06 in the
following discussion as HSE-1/4 functional.

As a next step, we used the so-called G$_0$W$_0$ approximation to
calculate the quasiparticle spectra, as implemented in
VASP.\cite{Shishkin1, Shishkin2, Fuchs, Shishkin3} This means that we
have used in G$_0$ and W$_0$ the Kohn-Sham eigenvalues and
orbitals. For W we take W$_0=\epsilon^{-1}V$, where the dielectric
matrix $\epsilon^{-1}_{{\bf G},{\bf G'}}({\bf q},\omega)$, with ${\bf
  G}$ and ${\bf G}'$ denoting reciprocal lattice vectors, is calculated
in the random phase approximation and the self-energy corrections are
evaluated to first order in the difference between the self-energy
$\Sigma$ and the Kohn-Sham potential.\cite{Louie,*Sham}. We note that
we also carried out calculations where G was updated by $n$ iterations
together with the DFT wave functions following the procedure described
in Ref.~\onlinecite{Shishkin3}. We denote this method by G$_n$W$_0$. 

In order to calculate the optical spectra, 

a) as a first approximation, we use the independent particle
approximation, i.e., using the energies and wavefunctions obtained in
LDA, GGA, HSE-1/4 and HSE-1/8.

b) Then, we use the GW approximation on top of the HSE-1/4 
and HSE-1/8 approximation. This corrects the quasiparticle
properties, such as the density of the states. A small correction due
to the GW relative to the starting point indicates that the starting
level of approximation may be good.

c) However, transition metal oxides contain electrons near the Fermi
level (in our case the the electrons occupying the Vanadium
$d$-states) which are expected to be strongly correlated. This means
that the electron-hole interactions are expected to be strong which
should play a significant role in excitonic particle/hole states.
These states are optically excited, therefore, we should and we will
include the role of the particle/hole interaction in the calculation
of the dielectric function. This is done using the Bethe Salpeter
equation (BSE).\cite{BSE}

\subsection{HSE-1/4 and HSE-1/8}

In this subsection we discuss the results for the single particle DOS
obtained with the two hybrid functionals HSE-1/4 and HSE-1/8 and their
GW corrections. The DOS as a result of the GW calculation 
is obtained by finding the average energy shift (relative to the Fermi energy)
for each quasiparticle energy level, and shifting the DOS calculated within DFT by
the corresponding amount. This method is used due to the fact that a GW calculation
with a really fine $k$-space mesh is extremely computationally expensive.
\begin{figure}
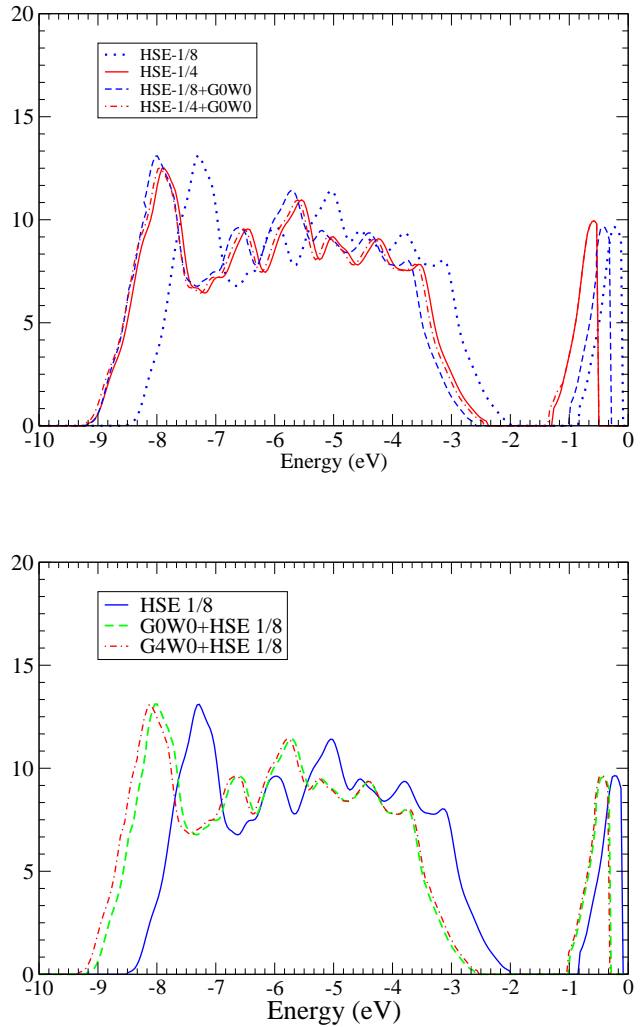

\vskip 0.3 in
\epsfig{file=Fig2a.eps,width=\figwidth} \\
\vskip 0.4 in
\epsfig{file=Fig2b.eps,width=\figwidth}
\caption{(Color online) Top:Results for the density of states (DOS) obtained
  with HSE-1/4 (red solid line) and HSE-1/8 (blue dotted line) and
  with HSE-1/4+G$_0$W$_0$ (red dot-dashed line) and HSE-1/8+G$_0$W$_0$
  (blue dashed line.) Bottom: Convergence study of the HSE-1/8+GW.
The solid blue line is the result obtained with just HSE-1/8. The green
dashed line is obtained by adding G$_0$W$_0$ to the HSE-1/8 and
the red dot-dashed line is obtained after iterating GW for the 4th
time starting from HSE-1/8. }
\label{figure2}
\vskip 0.3 in
\end{figure}

As can be observed in Fig.~\ref{figure2}, the results for the DOS
obtained using the HSE-1/4 (solid red line) functional and those
obtained using the HSE-1/8 functional are different, especially for
the Vanadium $d$-states just below the Fermi energy. The inclusion of
the G$_0$W$_0$ corrections has a small effect on the results of the
HSE-1/4 calculation and a much larger effect on the results obtained
starting from the HSE-1/8 hybrid. Notice, however, that the final
results, i.e., HSE-1/4+G$_0$W$_0$ and HSE-1/8+G$_0$W$_0$ are much
closer to each other, especially for the deeper (relative to the Fermi
energy) states, than the corresponding results without the G$_0$W$_0$
corrections.

In principle, if we were to carry out a perturbation series
calculation to all orders on top of either HSE-1/4 or HSE-1/8, the
final results should be the same.  We have carried out higher order GW
corrections, up to G$_4$W$_0$, on top of both HSE-1/4 and HSE-1/8. As expected
the results, obtained by iterating the GW four times on top of the
HSE-1/8, which is illustrated 
in Fig.~\ref{figure2}(bottom), show that the corrections
beyond the G$_0$W$_0$ are small. 
Therefore, this leads us to conclude that 
the remaining difference between the HSE-1/8+GW
and HSE-1/4+GW is not due to insufficient convergence of the GW 
iteration procedure. We think that this difference is due to processes not
captured by the GW approximation. As we will demonstrate 
here by solving the BSE equation, the effects of the quasielectron-quasihole
interaction are large. For example, the reason for the 
disagreement between HSE-1/8+GW and HSE-1/4+GW may be the fact that GW does 
not include the role of  virtual two-particle/one-hole excitations.
Furthermore, because the G$_0$W$_0$
corrections on top of the HSE-1/4 are smaller than the G$_0$W$_0$
corrections on top of the HSE-1/8, we may conclude that the HSE-1/4
provides a better starting point.

\subsection{Convergence of BSE Calculations}

The BSE calculation is computationally demanding because of the fact
that one has to include the interaction of particle/hole pairs into 
different relative momenta and from different bands.  As a result, to
make the calculation feasible within realistic computational time
scales, we need to limit the $k$-point size.

\begin{figure}
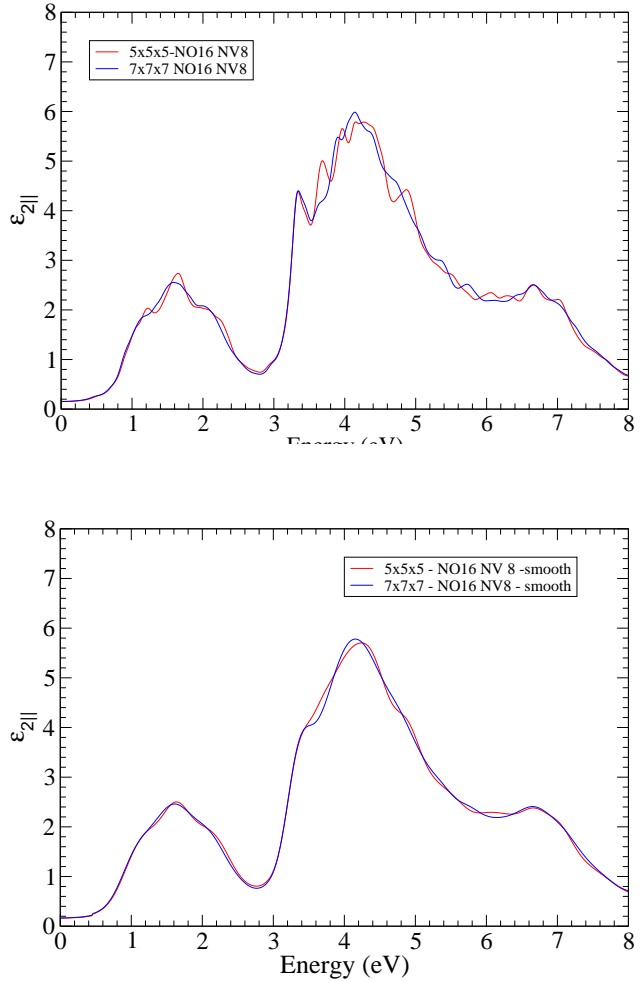

\vskip 0.3 in
\epsfig{file=Fig3a.eps,width=\figwidth} \\
\vskip .3 in
\epsfig{file=Fig3b.eps,width=\figwidth}
\caption{(Color online) (Top) The parallel to the a-axis imaginary
  part of the dielectric function as calculated for a $5\times 5\times
  5$ and a $7\times 7\times 7$ $k$-point set, by including the same
  number of occupied and unoccupied bands (18 occupied and 8
  unoccupied).  (Bottom) The data shown in the top part of this figure
  are smoothed using an exponentially weighted moving average
  technique. See Fig.~\ref{figure1} for the convention of
  parallel direction.}
\label{figure3}
\vskip 0.3 in
\end{figure}

We have studied the convergence with respect to the size of $k$-point
set used in our calculations. In Fig.~\ref{figure3} (top) we compare our
calculated imaginary part of the dielectric function $\epsilon_2$
parallel to the a-axis ($\epsilon_{2 ||}$) as calculated starting from
the HSE-1/4 and adding the G$_0$W$_0$ correction and solving the BSE
for a $5\times 5\times 5$ and a $7\times 7\times 7$ $k$-point
set,\cite{MP76} by including the same number of occupied and
unoccupied bands in both calculations. All of our results are smoothed
using a exponentially weighted moving average approach. This approach
leads to the curves shown in Fig.~\ref{figure3} (bottom). Notice that
the results of these two calculations agree reasonably well and, thus,
we have adopted the $5\times 5\times 5$ $k$-point set in our
calculations.

\begin{figure}
\vskip 0.3 in
\epsfig{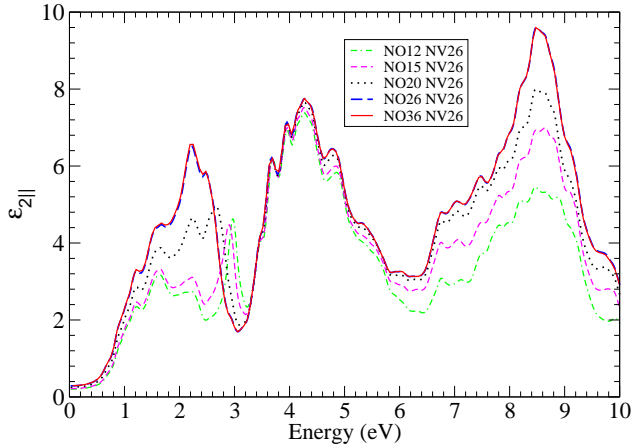}
\caption{(Color online) Convergence of $\epsilon_{2 ||}$ with
  number of occupied bands using a fixed number of unoccupied bands
  for the case of the full BSE calculation. See
  Fig.~\ref{figure1} for the convention of parallel direction.}
\label{figure4}
\vskip 0.3 in
\end{figure}

In Fig.~\ref{figure4} we demonstrate that using 26 occupied bands in our
calculation maybe sufficient to achieve a satisfactory level of
accuracy for $\epsilon_{2 ||}$.  In all the calculations presented in
Fig.~\ref{figure4}, we have used 26 unoccupied bands ($NV=26$) and we
varied the number of occupied bands, namely, we used
$NO=12,15,20,26,36$. Notice that the results for $NO=26$ have
virtually converged, namely, these results are very close to those
obtained for $NO=36$. We have also studied the same convergence 
with respect to the number of the unoccupied bands by keeping the
number of occupied bands fixed and found that when $NV=26$, the
results have converged. Therefore, all the results presented in this
paper were obtained with $NO=NV=26$.

\section{ Density of States}
\label{sec:dos}
\subsection{Comparison with photo-emission and band-structure}

In this section we present and compare the results obtained with
the various levels of approximation, namely, LDA, PBE, HSE, and HSE+GW.
In addition, we will compare with the photo-emission experiments.

\begin{figure}
\vskip 0.3 in
\epsfig{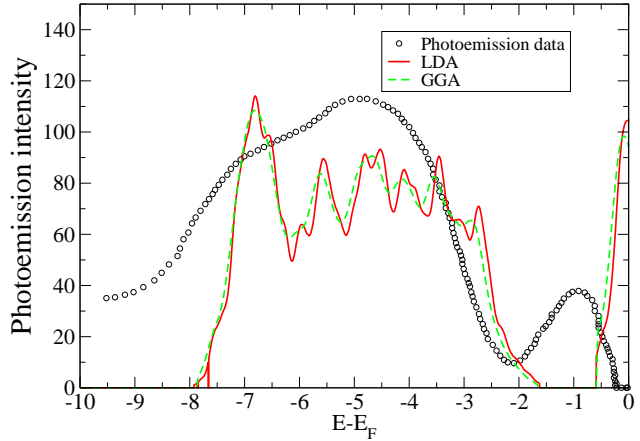} \\
\caption{The photoemission data taken from Ref.~\onlinecite{PES} is
  compared with the DOS of the occupied states as obtained from LDA,
  PBE calculations.}
\label{figure5}
\vskip 0.3 in
\end{figure}

\begin{figure}
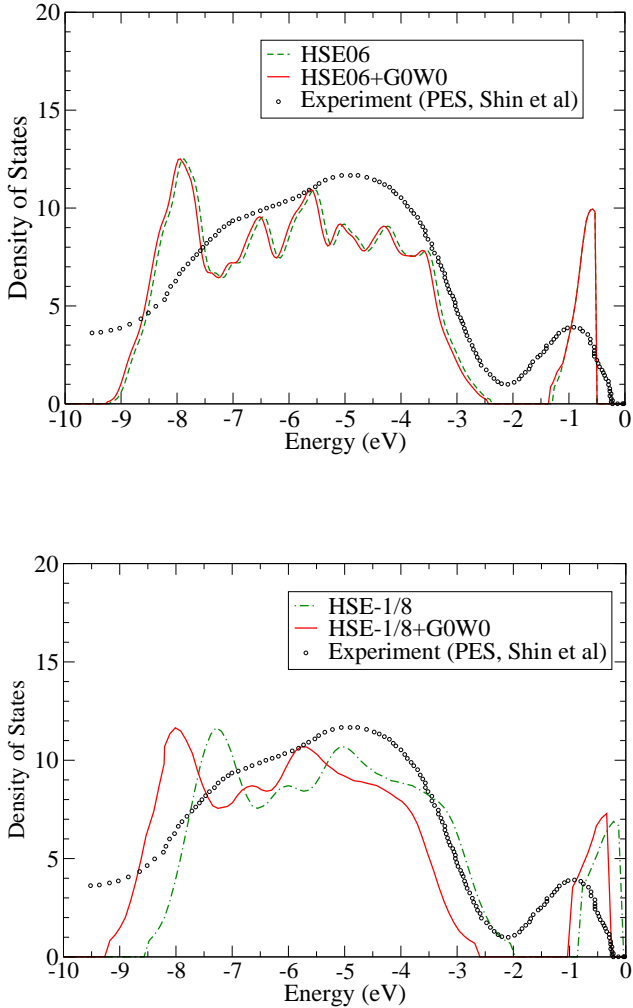

\vskip 0.3 in
\epsfig{file=Fig6a.eps,width=\figwidth} \\
\vskip 0.5 in
\epsfig{file=Fig6b.eps,width=\figwidth}
\caption{The photoemission data taken from Ref.~\onlinecite{PES} is
  compared with the DOS of the occupied states as obtained from
  HSE-1/4 and HSE-1/4+G$_0$W$_0$ (top) and HSE-1/8 and
  HSE-1/8+G$_0$W$_0$ calculations. Notice that the  band gap 
obtained is about 1.0 eV with HSE-1/4 and about 0.3 eV with HSE-1/8. }
\label{figure6}
\vskip 0.3 in
\end{figure}

In Figs.~\ref{figure5}, \ref{figure6} we compare the density of states
(DOS) as calculated using LDA, PBE, HSE-1/4 and HSE-1/8 with
photoemission data.\cite{PES} Notice that while the LDA and PBE
calculations give a reasonably good account for the occupied states
which are in the range of 2~eV to 8~eV below the Fermi level, both
fail to describe the states which lie just below the Fermi level. On
the contrary, it appears that the HSE-1/4 calculation gives a
reasonable account of the density of states in this narrow
region. This is a very interesting observation, which provides hope
that a perturbative scheme involving the GW approach and the BSE
starting from the HSE-1/4 wave functions might be a good idea.

\subsection{Band Structure}

\begin{figure}
\vskip 0.3 in
\epsfig{file=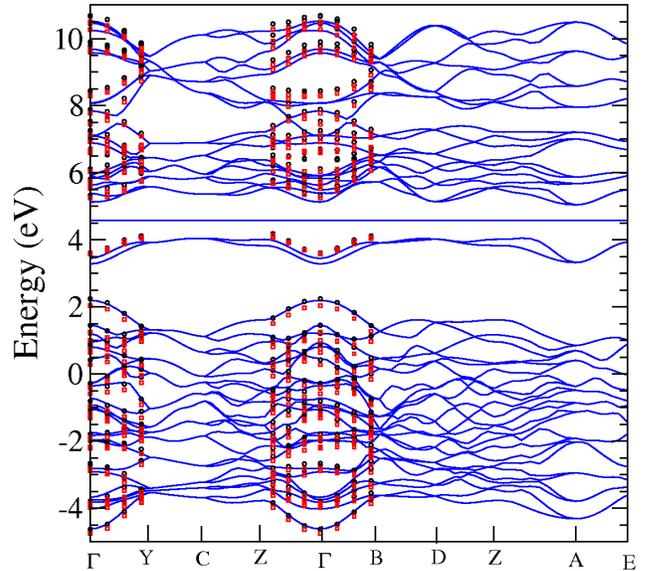,width=\figwidth}
\caption{(Color online) The band structure obtained with HSE-1/4
  (Solid lines) is compared to that obtained by HSE-1/4+G$_0$W$_0$
  (red squares) and HSE-1/4+G$_3$W$_0$ (black circles).  The results
  of the GW calculations have been shifted downward by significant
  constant amounts, namely, the G$_0$W$_0$ calculation by 0.98~eV and
  the G$_3$W$_0$ calculation by 0.60~eV to make their Fermi energies
  (5.50~eV and 5.18~eV respectively) the same with the Fermi energy of
  the HSE-1/4 calculation (4.58~eV).  Except for this overall
  constant, the bands from these calculations agree reasonably well
  with the HSE-1/4 calculation.}
\label{figure7}
\vskip 0.3 in
\end{figure}

This hope is also justified by comparing the results of the band
structure, obtained with HSE-1/4 with those obtained by GW based
calculations. In Fig.~\ref{figure7}, the results of G$_0$W$_0$
calculation on top of HSE-1/4 (solid blue lines) are shown as red
squares. In addition, we show the results of G$_3$W$_0$ calculation
with black circles. The overall energy scale has been changed by a
constant as explained in the figure caption, in order for the Fermi
levels to coincide. The overall small correction produced by GW
indicates that the perturbative corrections on top of HSE-1/4 are
small and, thus, we are hopeful that the perturbative series converges
fast.

\subsection{Partial density of states}
\label{PDOS}
In this subsection we discuss the changes that occur due to the
phase transition from the metallic rutile phase to the
insulating low temperature monoclinic M1 phase.

\begin{figure}
\vskip 0.3 in
\epsfig{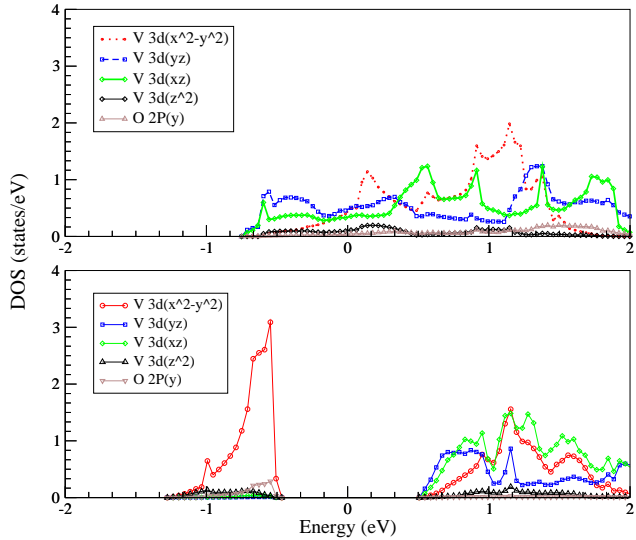} \\
\caption{The partial density of states near the Fermi level projected around
$V1$ for a $V$ state and around $O1$ for a $O$ 
state (see Fig~\ref{figure1}). 
Top: The rutile (metallic) phase. Bottom: The insulating (M1) low
temperature phase. In both cases we have kept only the orbitals 
which contain more than 0.1 electrons (out of the 4 electrons
total) below the Fermi level.}
\label{figure8}
\vskip 0.3 in
\end{figure}

In Fig.~\ref{figure8} we present the calculated partial density of states 
projected around
$V1$ and $O1$ (see Fig~\ref{figure1})  
for the rutile metallic phase (top) and the insulating M1 phase (bottom)
near the Fermi level. If we choose different atoms, clearly, the contribution
to the partial density of states of different orbitals will be changed
according to the relative position of the other atom.
 Notice that the states with the largest weight crossing 
the Fermi energy in the metallic phase are of the $t_{2g}$ symmetry. With our
choice of axes (the x-axis is along the rutile c-axis, the y-axis is 
along a Vanadium-Oxide direction in one of the octahedra
and the z-axis is along another Vanadium-Oxide direction 
in a different octahedron), the
orbitals with $t_{2g}$ symmetry are $d_{x^2-y^2}$, $d_{yz}$ and $d_{xy}$, while
the $d_{z^2}$ and $d_{xz}$ orbitals have $e_{g}$ symmetry.

In the insulating phase, the states contributing the most in the
total DOS just above the gap are
mainly of the same $t_{2g}$ symmetry. Just below the gap, however, only the
$d_{x^2-y^2}$ orbital gives most of the contribution to the density of
states. Namely, the dimerization of the V-V bond 
splits these $t_{2g}$ states in opposite directions around the
Fermi energy. The most affected orbital is the bonding combination
of the $d_{x^2-y^2}$ orbital, one lobe of which is directed along the V-V 
dimerization direction. Its energy is lowered relative to the Fermi level
due to the dimerization.
There is only one electron per V atom occupying 
these $d$ $t_{2g}$ states which are near and below the Fermi energy. 
By entering the dimerized phase, where the unit cell doubles
as shown in Fig.~\ref{figure1}, each 
of the two pairs of $d_{x^2-y^2}$ orbitals 
(one from each of the four V atoms in the unit cell) form
one bonding and one antibonding linear combinations which lead
to two bonding and two antibonding bands. 
The energy of these two bonding bands is separated by
a gap from the two antibonding bands and from the bands formed by all the other 
$t_{2g}$ states. Thus, the four electrons which correspond to the 
two pairs of vanadium atoms which participate in the formation of the 
two bonding
states, occupy and fill up all the states in the two bonding bands. 
These correspond to the two bands just below
the Fermi level shown in Fig.~\ref{figure7}.
This gives rise to an insulating state with the 
rest of the $t_{2g}$ states (along with the antibonding $d_{x^2-y^2}$ 
combinations) contributing to the bands just above the gap.

By inspecting Fig.~\ref{figure7}, it becomes clear that these bands are
more or less flat (a sign of one-dimensionality due to the
directional character of the dimerization) except along two 
directions. The direction 
$\Gamma \to B$ and the direction $A \to E$. The former direction is
along the dimerization direction (the a-axis in the
M1 phase, see Fig.~\ref{figure1}). 
The latter direction is along the b-axis of the M1 phase
(Fig.~\ref{figure1}). The former may be due to direct overlap
of the $d_{x^2-y^2}$ orbitals along the a-axis, while the latter
may be due to the hybridization of the $d_{x^2-y^2}$ with the
oxygen $p_y$. We find that the $V1$ and $V2$ atoms connected to $O1$
have approximately the same distance from $O1$ and the line that 
begins from $O1$ and bisects the $V1-V2$ line-segment is our $y$-axis.
As a result, we should expect to find the overlap between the 
$p_z$ (and the $p_x$) orbital of the $O1$ atom and the sum of the 
$d_{x^2-y^2}$ orbitals of the $V1$, and  $V2$ atoms  to
should be small. This is what we see in 
Fig.~\ref{figure8}, where the most significant contribution
to the PDOS just below the gap in the M1 phase is due to V $d_{x^2-y^2}$ and
the O $p_y$. This is not the case for the $O1'$ atom; in this case the
$V1$ and $V2$ atoms connected to $O1'$ are not equidistant from $O1'$ due to
the dimerization. As a result, the PDOS for atom $O1'$ shows contributions
from $p_x$ and $p_y$ orbitals.
 
\section{Dielectric Functions}
\label{sec:dielfunc}

Here we present our calculated imaginary part of the
dielectric function for momentum $q=0$, as a function of frequency (energy)
for light polarization parallel and perpendicular to the c-axis of the crystal.
This physical quantity is directly accessible by optical
studies.\cite{Epsilon2data} 

In order to calculate the dielectric function, which for zero momentum
transfer is directly proportional to the optical response, we will
first work within the independent particle approximation. In this case
the dielectric function is calculated using the energy eigenvalues
and eigenvectors as obtained by LDA or PBE or HSE and the 
non-interacting ``bubble-diagram'' for the response function.

\begin{figure*}
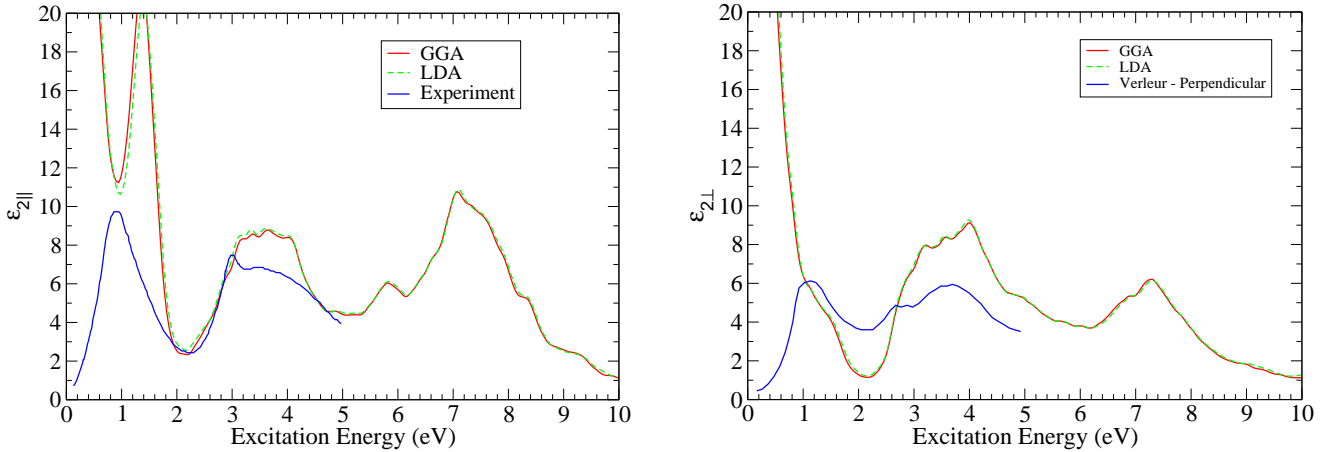

\vskip 0.3 in
\epsfig{file=Fig9a.eps,width=\figwidth} \qquad
\epsfig{file=Fig9b.eps,width=\figwidth}
\caption{(Color online) The experimental data taken from
  Ref.~\onlinecite{Epsilon2data} for the imaginary part of the
  dielectric function at zero momentum as a function of frequency for
  the cases of polarization parallel (left) and perpendicular (right)
  to the a-axis is compared with the results of LDA and PBE. See
  Fig.~\ref{figure1} for the convention of parallel direction.}
\label{figure9}
\vskip 0.3 in
\end{figure*}

In Fig.~\ref{figure9} the $q=0$ dielectric function as calculated
using LDA and PBE is compared to the experimental results.
Notice that while the higher energy peak is in reasonable
agreement with the experiment, this calculation misses entirely the
lower energy peak near $\sim 1$~eV and the overall distribution of strength of
the response function.

The electron-electron interaction, however, affects the excited
electron-hole pair, and it can create excitonic effects. In the
simplest extension, we can take into account these interaction effects
by solving the Bethe-Salpeter equation.\cite{BSE} In doing so, the
two-particle nature of the the BSE equation, makes the calculation
cumbersome, because the four point Green's function has to be solved
self-consistently. This calculation requires very large memory,
especially when one tries to include many occupied and unoccupied
bands and a larger $k$-point set, and a lot of CPU time.

\begin{figure*}
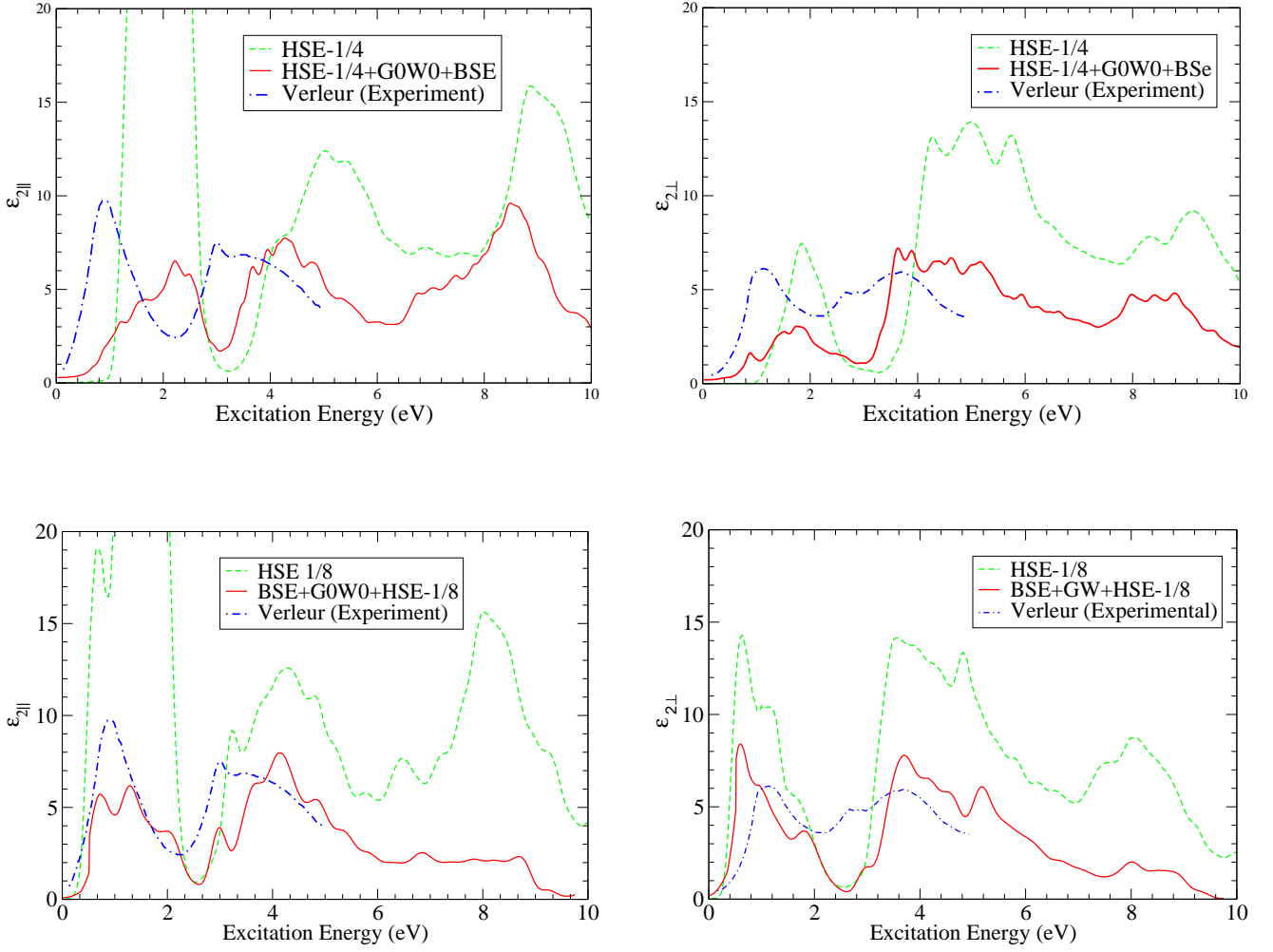

\vskip 0.3 in
\epsfig{file=Fig10a.eps,width=\figwidth} \qquad
\epsfig{file=Fig10b.eps,width=\figwidth} \\
\vskip 0.5 in
\epsfig{file=Fig10c.eps,width=\figwidth} \qquad
\epsfig{file=Fig10d.eps,width=\figwidth}
\caption{(Color online) The experimental data taken from
  Ref.~\onlinecite{Epsilon2data} for the imaginary part of the dielectric
  function for polarization parallel (left column) and perpendicular
  (right column) to the a-axis at zero momentum as a function of
  frequency, is compared with the results of HSE-1/4 and
  HSE-1/4+GW+BSE (top row) and with the results of the HSE1/8 and
  HSE1/8+GW+BSE (bottom row). See Fig.~\ref{figure1} for the
  convention of parallel direction.}
\label{figure10}
\vskip 0.3 in
\end{figure*}

We found that the first allowed transition appears at about  $\sim0.4~eV$ 
and
$\sim0.2~eV$   for    HSE-1/4+GW+BSE   and   HSE-1/8+GW+BSE   methods,
respectively,  which can  be  tentatively compared  to the  calculated
single  particle gaps  of  1.0~eV and  0.4~eV.  This is  a very  large
reduction (particularly  for the HSE-1/4 functional)  which implies strong
electron-hole interactions. We emphasize  here that very deep occupied
states  could seriously  influence  this value  indicating the  strong
correlation between the holes and  electrons. It may be suspected that
this  extremely  large  electron-hole  interaction  may  be  not  well
accounted for by the  perturbative BSE approximation in this  case. In the
top  of Fig.~\ref{figure10}  the results  for  the imaginary  part of  the
dielectric  function $\epsilon(q=0,\omega)$ for  polarization parallel
to the  crystalline a-axis are presented  for the two  cases where (a)
the HSE-1/4 and HSE-1/4+GW+BSE  and (b) HSE-1/8 and HSE-1/8+GW+BSE are
implemented   and   they    are   compared   with   the   experimental
data.\cite{Epsilon2data} In the bottom of Fig.~\ref{figure10} we present the
results  of  the  same  calculations  for  the  case  of  polarization
perpendicular to  the a-axis.  Notice  that the calculations  based on
the HSE-1/4 functional  show two main peaks at  approximately 2~eV and
4~eV which correspond to  the experimental peaks at approximately 1~eV
and 3~eV. Therefore, if the results of the full HSE-1/4+G$_0$W$_0$+BSE
are shifted  by a uniform  amount of about  1~eV at lower  energy they
would agree  with the  experimental position of  the peaks.   The main
reason for this disagreement arises from the HSE-1/4 calculation which
overestimates the  position of these  peaks by approximately  the same
amount. Notice, however, that the BSE calculation brings the intensity
of these peaks down to be comparable with the experimental intensity.  
Now, the results (right parts of Fig.~\ref{figure10}) obtained by reducing 
the mixing parameter $a$ in these calculations from its value of $a=1/4$
in the HSE-1/4 functional to $a=1/8$, show a better agreement with the
experimental  results.  This may  be  a  fortuitous coincidence  since
HSE-1/4 provides more accurate quasi-particle energies as confirmed by
GW  calculations. In  HSE-1/8  functional the  wavefunctions are  less
localized than  by HSE-1/4 functional which enter  BSE equations. That
might  compensate  the neglect  of  the  role  of the  electron-phonon
interaction in the imaginary part of dielectric function. Furthermore,
since we found  that the electron-hole interaction has  a large effect
on  the  dielectric function  $\epsilon(\omega)$,  one  would need  to
include   self-energy   corrections   due   to   two-particle+one-hole
intermediate states.

\subsection{Optical Transitions}
\begin{figure}
\vskip 0.1 in
\epsfig{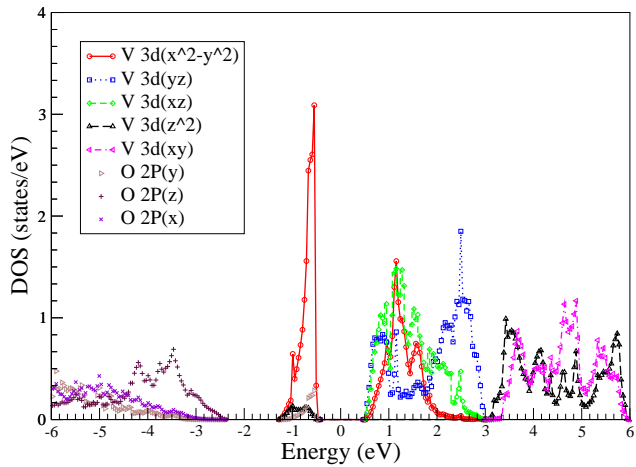}
\caption{(Color online) The PDOS projected as in Fig.~\ref{figure8}, 
expanded to show the nature of transitions observed in the optical spectra.
As before, states with small contribution are removed from the plot for 
clarity. }
\label{figure11}
\end{figure}

It is important to note that the nature of the optical transitions agrees 
well with previous analysis of the behaviour of this 
material \cite{PES,Eyert1}, as the
exact nature of these transitions has been in question for some time.
By comparing the optical spectra (Fig.~\ref{figure10}) to the expanded PDOS
in Fig.\ref{figure11}, we can see quite clearly the transitions occurring
at each peak in the spectra. It seems safe to do so using the projected
density of states within the HSE-1/4 picture, as the BSE calculation
did not largely shift the position of the peaks in the optical spectra, but
only modulated the intensity of the quasiparticle energies.  
The first peak in the optical spectra is 
entirely due to the excitation of the filled $t_{2g}$ states to the unfilled
anti-bonding $t_{2g}$ states. No other transitions at this energy range
are possible. At photon energies of $\sim 3eV$, 
the transition from filled $O$ $2p$ states to the same anti-bonding $t_2g$ 
states becomes available. The transition from the filled valence
band to the higher-energy $V$ $3d(e_g)$ anti-bonding states becomes
available at about $4 eV$, adding to the height of the 2nd peak. 
The third peak
at around $8 eV$ arises from the transition between $O$ $2p$ states and the
anti-bonding $V$ $e_g$ states. 

\section{Summary and conclusions}
\label{sec:summary}

We computed the ground state and optical properties of the M1 phase
of the VO$_2$ single crystal by means of 
density functional theory and beyond. We found that the HSE-1/4, i.e., the
HSE06 functional provides a better description to the  
single particle levels as inferred from quasi-particle
correction within GW approximation. However, the calculated gap is
about 1.0~eV which is about 0.6~eV 
larger than gap inferred by room-temperature photoemission 
measurements. We suspect that the neglected vertex
corrections in the quasi-particle correction and/or the electron-phonon
interaction are responsible for this discrepancy. 

In agreement with prior work, we find that the transition to the M1 phase
causes a Peierls-like distortion which, out of the three bands of $t_{2g}$ 
symmetry, it affects the band which is mainly of V $d_{x^2-y^2}$ character.
The four V atoms in the doubled unit cell after dimerization become two
pairs and their corresponding four $d_{x^2-y^2}$ states lead to two
bonding and two antibonding combinations separated by the Peierls gap.
There are four electrons available for these bands and after the 
dimerization, the two bonding bands just below the Fermi energy become
fully occupied, which leads to a Peierls insulator (for details see our
Sec.~\ref{PDOS}).

A major new
finding in our paper is that the electron-hole interaction in these
oxide materials is very strong which leads to significant reduction of
the overall response function relative to the starting hybrid
functional results. We believe that this may be due to the pronounced 
localized nature of the
$d_{x^2-y^2}$ orbital and the strong electron correlation
arising from the charge localization. 
Occupied electron states with energy 10~eV below
the Fermi-level have \emph{substantial} effect on the calculated
first absorption peak ($\sim$0.4~eV).  This gives us hope that
the screened Coulomb interaction between electrons and 
holes, which is strong in the M1 phase of VO$_2$, might cause  
a high impact ionization rate, leading to multi-exciton generation 
upon high-energy
excitation. We believe that this finding could be rather general
among TMOs where the same interaction is responsible for
the Mott-gap in this class of materials. Thus, Mott-insulators could
indeed be an important class of materials of interest for photovoltaic
applications. In addition to opto-electronic excitations, 
doping and extraction of carriers should be addressed in future studies in
order to fully explore the applicability of this class of materials in
third generation solar cells.

On the other hand, our results imply that it is very difficult to
provide quantitatively accurate predictions for the optical response in
the complex materials such as the M1 phase of VO$_2$, even by means of
very sophisticated methods involving many-body perturbation theory.
The disagreement with the experiment on the location of the two main
(low energy) peaks in the imaginary part of the dielectric function
might be attributed to the fact that we have neglected vertex
corrections and the contribution of two-particle-one hole states
in the quasiparticle self-energy,
the electron-phonon interaction. Furthermore,  it might be 
significant the contribution to the response function from three-particle
three-hole states in this strongly correlated material which is
obviously not captured by BSE. Nevertheless, the HSE06+G$_0$W$_0$+BSE
results on the optical response in the energy region important for
photo-voltaic applications, show a very large improvement relative 
to that obtained by independent particle approximations where the
intensity of the response function by HSE06+G$_0$W$_0$+BSE is
in good agreement with the experiments. This finding indicates that
HSE06+G$_0$W$_0$+BSE method can be applied to semi-quantitatively
predict optical excitations in transition metal oxides.

\section{Acknowledgments}

This work was supported in part by the U.S.\ National High Magnetic 
Field Laboratory which is partially funded by the U.S.\ National 
Science Foundation.
 
\bibliography{mybib}

\end{document}